\documentstyle[11pt,newpasp,twoside]{article}
\input psfig.sty
\index{summary}
\index{instructions}
\index{template}

\newcommand\E[1]{\times10^{#1}}
\newcommand\U[1]{{\,\rm #1}}
\newcommand\rs[1]{_{#1}}

\reversemarginpar

\begin{document}
\title{Implications of the synchrotron emission in plerionic nebulae:
  a second component in the Crab Nebula}
 \author{Rino Bandiera}
\affil{Osservatorio Astrofisico di Arcetri, Largo E. Fermi 5, Firenze (Italy)}

\begin{abstract}
High-quality spatially resolved spectra of synchrotron emission from plerions
represent a powerful diagnostic tool for the physical conditions in these
objects.
A fundamental question is to what extent spatial variations of the synchrotron
spectrum originate from the evolution of a single population of injected
electrons, and when instead they imply the coexistence of different populations
of electrons.
Here I shall discuss observations of the Crab Nebula at millimetric wavelengths
which, compared with a radio map, show evidence for the emergence of a second
synchrotron component, characterized by a flatter spectrum and undetected in
the radio range.
This component is confined to the inner part of the nebula, and cannot
originate from synchrotron-dominated evolution, also because in the Crab Nebula
the electrons emitting at millimeter wavelengths evolve almost adiabatically.
Then its discovery indicates that there are more populations of injected
particles.
Finally, the detection of a low-frequency break in the spectra of synchrotron
filaments can be interpreted as an effect of higher magnetic fields in
filaments.
\end{abstract}

\section{Introduction}

Synchrotron emission from Crab-like nebulae (plerions) provides information on
magnetic fields and relativistic electrons contained in these nebulae.
One of the goals of the analysis of this emission is to determine the
characteristics of the particles injected in the nebula, and eventually those
of the injection processes.
However in order to obtain the injected spectrum it is necessary to correct for
the evolution of particles, which are subject both to adiabatic and synchrotron
losses: the former ones dominate at lower particle energies and preserve the
shape of the distribution, while the latter ones dominate for more energetic
particles, causing a steepening in their distribution.

In the integrated synchrotron spectrum at least two regions with different
spectral indices are expected, separated by a break at frequency $\nu\rs{b}$,
where the adiabatic and synchrotron losses are equal (for a more detailed
treatment see Pacini \& Salvati 1973).
At frequencies higher than $\nu\rs{b}$ spectral inhomogeneities in the spectral
index are generally explained in terms of synchrotron evolution; while below
$\nu\rs{b}$ the power-law index of the present particle distribution must
correspond to that at injection.
In the case of the Crab Nebula, where $\nu\rs{b}\sim15,000\U{GHz}$, Bietenholz
et al.\ (1997) show that the radio spectral index is highly homogeneous, with
an upper limit of only $0.01$: this result has been interpreted as a compelling
evidence in favour of a single component of injected particles.

Here I shall report on observations at mm wavelengths (a work in collaboration
with R.
Cesaroni and R. Neri), which instead indicate the coexistence of two
components.
Details on the observations as well as an extensive discussion of the
results are presented by Bandiera et al.\ (2001).

\section{230~GHz observations of the Crab Nebula}

\begin{figure}
\centerline{ \psfig{file=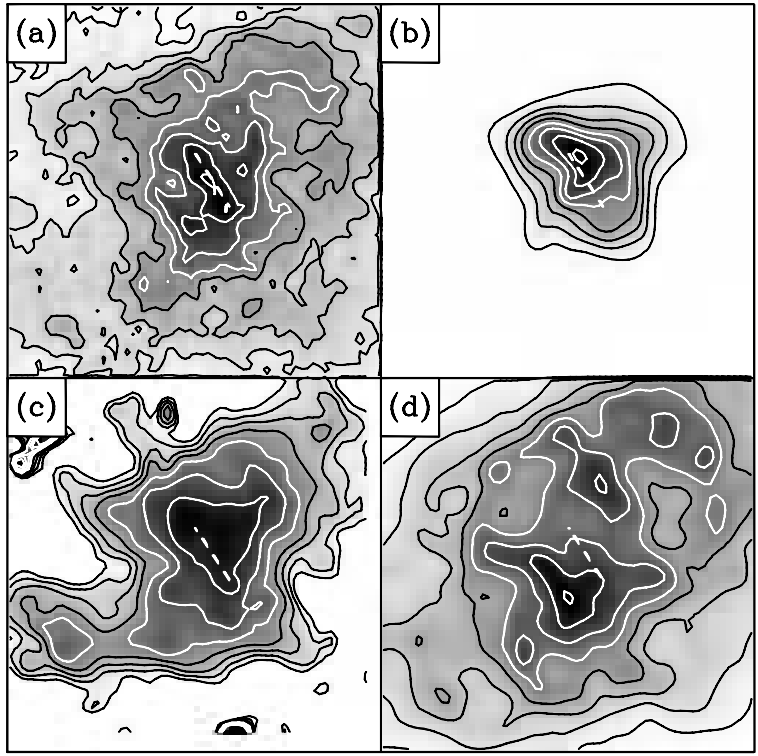,height=10cm,angle=-90} }
\vspace{4.0mm}
\caption{The map of the second component, as extracted from the mm map (a), is
well correlated in size, position and shape with maps at higher frequencies,
like the X-ray map (b) and the optical spectral map (c); it is instead
anticorrelated with the radio map (d).
}
\end{figure}

The Crab Nebula has been mapped at 230~GHz using the MPIfR bolometer array
MAMBO, at the 30-m IRAM telescope (Pico Veleta, Spain), with a $10\farcs5$
angular resolution and a 10\% photometric accuracy.
The 230~GHz map has been compared with a 1.4~GHz VLA map (kindly provided by
M.F. Bietenholz), suitably downgraded to the same resolution.
The patterns in the two maps are similar, but not identical.
This is apparent from the spectral map between the two frequencies (Bandiera et
al.\ 2001): this map shows that in the inner region the spectral index is
flatter (by $\sim 0.05$) than in the rest of the nebula, and is also flatter
than the radio spectral index (a similar effect is seen also at 850~$\mu$m;
Green, at this meeting).
This means that, in the range from radio to mm wavelengths, the spectrum of the
inner nebula is concave, a behavior hard to explain in terms of synchrotron
losses, which on the contrary generate steeper spectra at higher frequencies.
Moreover, since $\nu\rs{b}$ is much higher than 230~GHz, synchrotron losses are
not expected to affect the electron distribution in this energy range.
A concave spectrum, instead, naturally appears in the case of the coexistence
of two different power-law components.

With this in mind we have mapped the residuals after having subtracted an
extrapolation of the radio map (after PSF equalization).
The map of residuals shows an excess in the inner region, plus a pattern of
secondary negative features, whose positions coincide with those of radio
filaments.
By a non-linear filtering we have synthesized a map of the filaments and we
have combined it with the map of residuals in a way to minimize the ripples.
We came out with a map of what we guess is the second component (Fig.~1a).
In size, shape and position this component resembles the X-ray map (Fig.~1b)
and the optical flatter spectrum region (Fig.~1c; V\'eron-Cetty \& Woltjer
1993): this is an unexpected result, because the latter two maps were believed
to be shaped by synchrotron losses, while the mm component is not.
Finally, this second component is anticorrelated with the radio map (Fig.~1d).

\begin{figure}
\centerline{ \psfig{file=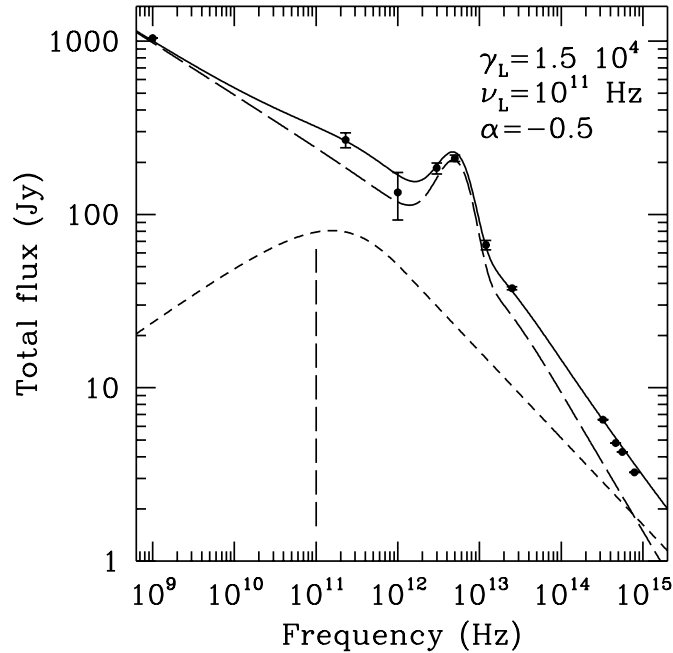,height=8.6cm,angle=0} }
\vspace{4.0mm}
\caption{Two-component spectral fit to photometric data from radio to optical
wavelengths.
The second component is represented by the short-dash line which, summed with
other component, accounts for the integrated spectrum (solid line).
The vertical dash line indicates the characteristic frequency associated with
$E\rs{L}$, the low cutoff energy.
}
\end{figure}

\section{Discussion}

Let me first discuss the negative features coincident with radio filaments.
They indicate that, in between radio and mm wavelengths, the spectrum of the
filaments is steeper than the integrated spectrum.
If this is ascribed to the presence of a synchrotron break, this break must be
located at about 70~GHz, namely at a frequency $\sim200$ times lower than the
$\nu\rs{b}$ for the integrated spectrum.
If electrons cannot diffuse through filaments we have $\nu\rs{b}\propto
B^{-3}$: thus in filaments the field should be $\sim6$ times higher than the
average.

The fact that even these secondary features are real makes us confident that
the primary feature in the residual map, namely the excess in the inner nebula,
is not just an instrumental artifact.
Even though the 230~GHz map has been compared with a radio map obtained about
10 years earlier, and even though there is evidence of rapid variability in
radio (Bietenholz et al.\ 2001; and at this meeting), there are various reasons
why the observed excess cannot be an artifact of time variability.
First of all, Bietenholz et al.\ show a wavy structure of the variability
pattern that, when smeared to $10\farcs5$ resolution, should average down to
zero.
Moreover the extension we find is larger than that of the time-variable region
in radio.
Finally if there have been ``bursts of injected particles'', their effect
should be detectable for a long time; on the contrary, the evolution of
the Crab integrated flux is found to be smooth (Aller \& Reynolds 1985).

An analysis of the integrated spectrum of the Crab Nebula allowed us to exclude
a thermal origin (either dust or free-free) for this component.
Even in the case of a synchrotron nature stringent limits on the electron
distribution follow from the data at other wavelengths.
The absence of this component at radio wavelengths implies a low-energy cutoff
at a $E\rs{L}\sim1.5\E{4}m\rs{e}c^2$; moreover, its emergence at optical
frequencies and beyond constrains the slope of the distribution above $E\rs{L}$
to be about --2 (about --0.5 in the spectrum; Fig.~2).

In this way a total number of $N\rs{tot}\sim2\E{48}$ electrons is estimated for
this component (for an assumed nebular field $\sim3\E{-4}\U{G}$).
This value for $N\rs{tot}$ is in nice agreement with what predicted by Kennel
\& Coroniti (1984) model, which otherwise fails to account for the number radio
emitting electrons.
This may indicate that the Kennel \& Coroniti mechanism is producing the newly
discovered component, while the radio component requires a different
mechanism.

\acknowledgements{This work has been supported by the Italian Ministry for
University and Research under Grant Cofin99--02--02, and by the National
Science Foundation under Grant No.\ PHY99-07949.}

\end{document}